\documentclass[journal,twocolumn,10pt]{IEEEtran}
\usepackage{graphicx}
\usepackage{verbatim}
\usepackage{upgreek}
\usepackage{amssymb,amsmath}
\usepackage{color}
\usepackage{epstopdf}
\usepackage{bm}
\usepackage{cite}
\usepackage{array,color}
\usepackage{algorithm}
\usepackage{algorithmicx}
\usepackage{algpseudocode}
\usepackage{amsmath}
\usepackage{amsfonts}
\usepackage{amsthm}
\usepackage{graphics}
\usepackage{epsfig}
\usepackage{soul}
\usepackage{booktabs}
\usepackage{multirow}
\usepackage{stfloats}
\soulregister\cite7
\soulregister\ref7
\usepackage{subfigure}
\usepackage{tabularx}
\usepackage{makecell}
\usepackage{graphicx}
\usepackage{gensymb}
\usepackage{caption}
\renewcommand\appendix{\setcounter{secnumdepth}{3}}
\newtheorem{lemma}{Lemma}

\usepackage{etoolbox}
\newcommand{\tabincell}[2]{\begin{tabular}{@{}#1@{}}#2\end{tabular}} 
\makeatletter
\patchcmd{\@makecaption}
{\scshape}
{}
{}
{}
\makeatletter
\patchcmd{\@makecaption}
{\\}
{.\ }
{}
{}
\makeatother

\ifCLASSINFOpdf
\else
\fi
\begin{document}

\title{\huge Towards Intelligent Edge Sensing for ISCC Network: Joint Multi-Tier DNN Partitioning and Beamforming Design}

\author{Peng Liu, Zesong Fei,~\IEEEmembership{Senior~Member,~IEEE},  Xinyi Wang,~\IEEEmembership{Member,~IEEE}, \\  Xiaoyang Li, \IEEEmembership{Member,~IEEE}, Weijie Yuan,~\IEEEmembership{Senior~Member,~IEEE},  Yuanhao Li, \IEEEmembership{Senior~Member,~IEEE}, \\ Cheng Hu,~\IEEEmembership{Senior~Member,~IEEE}, and Dusit Niyato,~\IEEEmembership{Fellow,~IEEE}
	
	\thanks{Peng Liu, Zesong Fei, and Xinyi Wang are with the School of Information and Electronics, Beijing Institute of Technology, Beijing 100081, China (e-mail: bit\_peng\_liu@163.com, feizesong@bit.edu.cn, bit\_wangxy@163.com).}
		\thanks{Xiaoyang Li is  with Shenzhen Research Institute of Big Data, The Chinese University of Hong Kong (Shenzhen), Shenzhen 518172, China (e-mail: lixiaoyang@sribd.cn).}
		\thanks{Weijie Yuan is with  Department of Electronic and Electrical Engineering, Southern University of Science and Technology, Shenzhen 518055, China (e-mail: yuanwj@sustech.edu.cn).}
	\thanks{Yuanhao Li and Cheng Hu are with the Radar Technology Research Institute, School of Information and Electronics, Beijing Institute of Technology, Beijing 100081, China(e-mail: lyh.900101@163.com, cchchb@163.com).}
	\thanks{Dusit Niyato is with the College of Computing and Data Science, Nanyang Technological University, Singapore (e-mail: dniyato@ntu.edu.sg).}
}

\maketitle

\begin{abstract}
The combination of Integrated Sensing and Communication (ISAC) and Mobile Edge Computing (MEC) enables devices to simultaneously sense the environment and offload data to the base stations (BS) for intelligent processing, thereby reducing local computational burdens. However, transmitting raw sensing data \textcolor{black}{from ISAC devices to the BS} often incurs substantial fronthaul overhead and latency. This paper investigates a three-tier collaborative inference framework enabled by Integrated Sensing, Communication, and Computing (ISCC), where cloud servers, MEC servers, and ISAC devices cooperatively execute different segments of a pre-trained deep neural network (DNN) for intelligent sensing. By offloading intermediate DNN features, the proposed framework can significantly reduce fronthaul transmission load. Furthermore, multiple-input multiple-output \textcolor{black}{(MIMO)} technology is employed to enhance both sensing quality and offloading efficiency. To minimize the overall sensing task inference latency across all ISAC devices, we jointly optimize the DNN partitioning strategy, ISAC beamforming, and computational resource allocation at the MEC servers and devices, subject to sensing beampattern constraints. \textcolor{black}{We also propose an} efficient two-layer optimization algorithm. In the inner layer, \textcolor{black}{we derive} closed-form solutions for computational resource allocation using the Karush-Kuhn-Tucker conditions. \textcolor{black}{Moreover, we design} the ISAC beamforming vectors via an iterative method based on the majorization–minimization and weighted minimum mean square error  techniques. In the outer layer, \textcolor{black}{we develop a} cross-entropy-based probabilistic learning algorithm to determine an optimal DNN partitioning strategy. Simulation results demonstrate that the proposed framework substantially outperforms existing two-tier schemes in inference latency.

\vspace{2ex}
\textbf{Keywords: Integrated sensing and communication, mobile edge computing, DNN partitioning,  multiple-input multiple-output, cross-entropy.} 
\end{abstract}

\IEEEpeerreviewmaketitle
\section{Introduction}
\textcolor{black}{Burgeoning applications, including autonomous driving, unmanned aerial vehicles (UAVs), extended reality,} and industrial automation, demand not only ultra-reliable and low-latency communication, but also real-time and high-resolution environmental sensing.  These dual demands challenge conventional wireless systems, where the separation of sensing and communication often results in inefficient spectrum usage and redundant hardware deployment. Integrated Sensing and Communication (ISAC) has emerged as a promising paradigm to overcome these limitations by enabling joint design and operation of sensing and communication within a unified framework \cite{2022Liu}. By sharing spectral and hardware  resources, ISAC enhances spectral efficiency while reducing system cost.  To fully unlock the potential of ISAC in 6G networks, extensive research has been devoted to key enabling technologies, particularly in waveform design~\cite{Liu2018tsp, Wang2022tvt}, beamforming optimization~\cite{liux2020tsp, beam1, Yang2025tsp}, multiple access~\cite{access, access2}, and resource allocation strategies~\cite{yangtwc2024, Yu2023tvt}.

Despite its numerous advantages, the sensing process in ISAC systems typically generates a large volume of raw data to be processed for target recognition and classification. These applications typically require intensive computation and substantial processing power, which often exceeds the capabilities of devices with limited energy and computational resources. To address this challenge, mobile edge computing (MEC) has been introduced as a complementary technology that enables computation offloading to nearby edge servers offering higher processing capacity and low latency for intelligent processing \cite{SunMEC}. The integration of MEC with ISAC not only enables low-latency processing of sensing data but also facilitates the realization of edge intelligence \cite{Xu2023,EdgeAIwen,Lu2024}. This convergence has led to the emerging paradigm of Integrated Sensing, Communication, and Computing (ISCC) \cite{Wensurvey2024}, which aims to support the growing demand for real-time, data-driven, and computation-intensive intelligent applications in 6G networks.

Recent research has explored the potential of ISCC from the perspectives of system architecture design and performance optimization \cite{ISCC2, liu2024joint,Pengtvt2,huang2024,aircomp1,aircomp2,aircomp3, ISCC5,ISCC6 }. In \cite{ISCC2}, the authors proposed an ISCC framework in which terminals simultaneously perform sensing and computation offloading over shared spectral resources. By jointly optimizing beamforming vectors and computational resource allocation, the energy consumption of terminals was significantly reduced. Building on this, the authors in \cite{liu2024joint, Pengtvt2} developed a three-tier ISCC architecture comprising a cloud server, multiple MEC servers, and several devices, where ISAC beamforming matrices and offloading strategies were jointly optimized to minimize the average execution latency of sensing tasks. In \cite{huang2024}, a short-packet transmission-based ISCC paradigm was introduced, where ISAC devices transmit sensing data to edge servers, with the packet size and computational resources jointly optimized to effectively reduce device energy consumption.  Li \textit{et al.} proposed a novel ISCC framework integrated with over-the-air computation (AirComp), which enables the functional computation of signals based on the analog wave superposition property of multiple access channels \cite{aircomp1,aircomp2}. Furthermore, the authors in \cite{aircomp3} investigated an AirComp-enabled ISCC system tailored for wireless sensor networks, where the beampatterns of the sensors are jointly designed to minimize AirComp errors.  Different from the aforementioned works where terminal devices are equipped with ISAC capabilities, in \cite{ISCC5}, a Non-Orthogonal Multiple Access (NOMA)-aided multi-tier ISCC architecture was proposed, where the base station (BS) performs sensing while concurrently supporting multiple users with edge computing, assisted by a cloud server. Furthermore, the authors in \cite{ISCC6} investigated the deployment of multiple cloud servers and an ISAC BS, and proposed a two-tier offloading strategy to minimize device energy consumption. 

Although prior studies have \textcolor{black}{investigated the offloading of sensing tasks to edge or cloud servers,} they typically treated the sensing task as a monolithic entity. This requires transmitting large volumes of raw sensing data, which will lead to significant communication delays and raise privacy concerns. To address these challenges, deep neural network (DNN) partitioning has emerged as a promising solution \cite{dnn1}. Since a typical DNN comprises multiple layers with millions of parameters, executing the entire model incurs high computational cost. Fortunately, the intermediate features generated by DNN layers are often significantly smaller than the raw input data, enabling feature-level offloading to effectively reduce transmission overhead \cite{dnn2}. By leveraging the hierarchical structure of DNNs, partitioning allows for flexible distribution of inferring workloads across devices, edge servers, and the cloud, thereby reducing latency, enhancing privacy, and improving overall sensing efficiency. Recently, several studies have investigated the integration of DNN partitioning into MEC systems. In \cite{dnnmec1}, the authors proposed a channel-adaptive partitioned inference strategy, where a U-shaped network is collaboratively executed on the MEC server and the devices, with the partition point dynamically adjusted according to wireless channel conditions.  Furthermore, the authors in \cite{dnnmec3} proposed a learning-and-optimization algorithm that jointly determines the DNN partitioning and resource allocation to reduce device energy consumption.

Regarding the integration of DNN partitioning with ISCC systems, the authors in \cite{dnniscc1} proposed a novel edge intelligence framework in which ISAC devices perform radar sensing, extract features from raw sensing data, and transmit them to a MEC server for further processing \textcolor{black}{in a time-division manner}. In particular, the device transmit power and communication time were jointly optimized to maximize inference accuracy. Building on this, the authors in \cite{dnniscc2} and \cite{wen2024aircomp} introduced AirComp, enabling the feature extracted from multiple devices to be superimposed over the air, followed by inference at the edge server's DNN. In \cite{dnniscc3}, a collaborative inference architecture involving UAVs and MEC servers was proposed, where user association, DNN partitioning, and beamforming vectors were jointly optimized under constraints on sensing beampattern gain, inference latency, and device energy consumption, aiming to maximize the total offloading rate. However, most of these works assumed a fixed partition point \cite{dnniscc1, dnniscc2} or adopted only two-tier partitioning between the device and MEC server \cite{dnniscc1, dnniscc2, wen2024aircomp, dnniscc3}, without fully leveraging the heterogeneous computational resources available across different tiers and network entities. Moreover, they typically considered single-antenna ISAC devices or MEC servers, overlooking the potential of multiple-input multiple-output (MIMO) technique in enhancing offloading efficiency and improving sensing performance. While integrating multi-layer DNN partitioning into MIMO-ISCC systems can offer increased flexibility and better latency performance, it also introduces considerable design complexity due to the intricate coupling among partitioning strategies, MIMO beamforming, and resource allocation. 

Motivated by the above analysis, we propose a three-tier cloud–edge–device  collaborative framework that integrates DNN partitioning and employs MIMO technique to enhance both sensing and offloading performance. Specifically, we consider an ISCC system comprising multiple multi-antenna devices that transmit ISAC signals for wireless sensing and perform intelligent signal processing via a pre-trained DNN model. To alleviate the computational burden and reduce inference latency, each device executes part of DNN locally and offloads intermediate features—rather than raw sensing data—to the base station (BS) via ISAC signals for further processing. Unlike existing works \cite{dnniscc1, dnniscc2, wen2024aircomp, dnniscc3}, which considered only two-tier DNN partitioning between devices and MEC servers, we introduce a three-tier collaborative architecture by incorporating a cloud server. Taking into account the computational capacities of the cloud, MEC server, and devices, as well as the channel conditions and sensing requirements, we jointly optimize the DNN partitioning strategy, ISAC beamforming vectors, and computational resource allocation to minimize the total inference latency of sensing tasks. The main contributions of this paper are summarized as follows:

\begin{itemize}
\item{We propose an ISCC-based three-tier collaborative inference framework, in which the cloud server, MEC server, and devices collaboratively execute different segments of a pre-trained DNN to accomplish intelligent sensing tasks. The devices transmit ISAC signals to simultaneously perform environmental sensing and offload intermediate DNN features to the MEC and cloud servers for further inference. Within this framework, we develop a comprehensive inference latency model that jointly considers ISAC signaling, DNN partitioning, and computation offloading.}

\item{We formulate a latency minimization problem through jointly optimizing the DNN partitioning strategy, computational resource allocation, and ISAC beamforming vectors, subject to sensing beampattern matching and the computational capacity constraints of MEC servers and devices. To address the resulting mixed-integer nonlinear programming (MINLP) problem, we exploit its hierarchical structure and decompose it into an inner-layer subproblem for computational resource allocation and ISAC beamforming, and an outer-layer subproblem for DNN partitioning.}

\item{For the inner problem, we derive closed-form solutions for both local and MEC resource allocation using the Karush-Kuhn-Tucker (KKT) conditions. Moreover, we develop an iterative algorithm based on the majorization–minimization (MM) framework, which incorporates the weighted minimum mean square error (WMMSE) method and the orthogonal Procrustes problem (OPP) transformation to obtain closed-form ISAC beamforming vectors. For the outer problem, a cross-entropy (CE)-based algorithm is introduced to iteratively \textcolor{black}{learn} the optimal distribution of the DNN partitioning strategy.}
\item{\textcolor{black}{We provide} extensive simulation results to validate the effectiveness of the proposed three-tier collaborative inference architecture. \textcolor{black}{In particular, a comparison with the benchmark branch-and-bound (BnB) algorithm shows that the proposed algorithm attains optimal DNN partitioning with significantly reduced computational complexity.} Moreover, compared with existing two-tier collaborative frameworks, the proposed approach achieves lower inference latency for sensing tasks. Furthermore, the results highlight a trade-off between sensing beamforming gain and inference latency.}

\end{itemize}

The remainder of this paper is organized as follows. Section II presents the system model, including the ISAC signal model and the computation model related to DNN partitioning, followed by the formulation of the joint optimization problem. In Section III, we propose a two-layer optimization framework that jointly optimizes the DNN partitioning strategy, ISAC beamforming, and computational resource allocation. Section IV provides simulation results to validate the effectiveness of the proposed architecture and algorithm. Finally, Section V concludes the paper.


\section{System Model}

\begin{figure}[!t]
	\centering
	\includegraphics[width=3.5in]{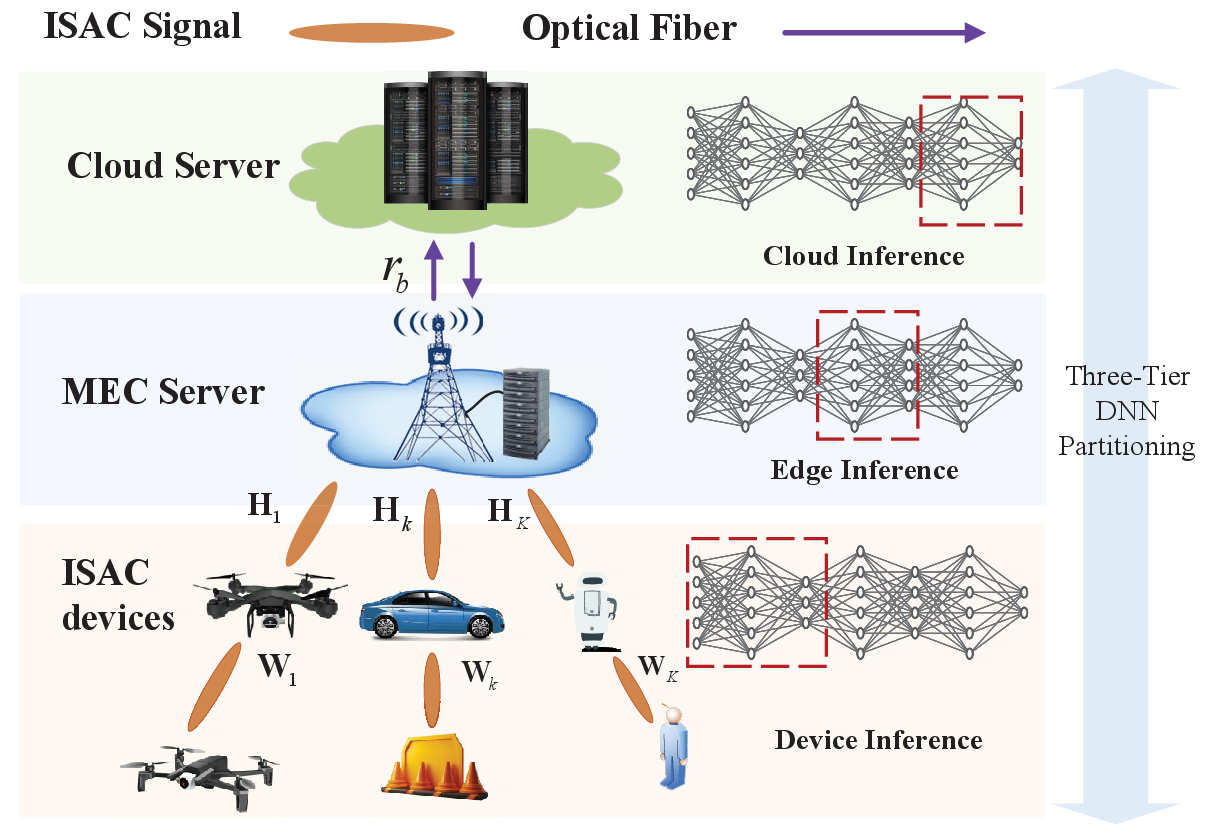}
	\caption{The ISAC-enabled cloud-edge-device three-tier network architecture.}
	\label{fig:sys}
\end{figure}
As shown in Fig. \ref{fig:sys}, we consider a three-tier cloud-edge-device  network architecture consisting of $K$ ISAC devices, denoted by $\mathcal{K}=\{1,\ldots,K\}$,  an $M$-antenna BS equipped with an MEC server, and a cloud server\footnote{\textcolor{black}{This model can also be applied to scenarios with multiple servers. In such cases, user association has to be carefully designed while considering both varying channel quality and computational load balancing across different servers, in order to maintain high offloading and execution efficiency. Due to the limited space in this paper, we leave this issue as future work.}}. Each ISAC device needs to perform a latency-sensitive and computation-intensive DNN-based intelligent sensing task, such as target classification and recognition. Specifically, each ISAC device first acquires raw data through wireless sensing, and then preprocesses the echo signals into target-related spectrogram images using data sampling, filtering, and short-time Fourier transform (STFT) \cite{stft}. Finally, a pre-trained DNN model is employed to classify targets such as pedestrians and vehicles. Due to the limited computational capability and energy constraints of ISAC devices, executing a large-scale DNN model locally can lead to high inference latency, while offloading large volumes of raw data to the edge server may introduce excessive transmission delay and potential security risks. Inspired by \cite{dnn1}, we adopt a multi-tier collaborative inference approach that integrates DNN partitioning and task offloading to address this challenge. Specifically, each ISAC device processes the early layers of a complex DNN model, while the MEC server and cloud server handle the remaining layers to generate final inference result. Since the data volume significantly decreases at some intermediate DNN layers, such as pooling layers, the partitioning strategy not only alleviates the computational burden on devices, but also reduces data offloading pressure and enhances raw data privacy protection. To improve spectral efficiency, we assume that each ISAC device  is equipped with $N_t$ transmit antennas and transmits ISAC signals to simultaneously perform task offloading and target sensing, and space division multiple access (SDMA) is employed to achieve greater bandwidth, thereby enhancing sensing accuracy and communication rates \cite{Liu2018tsp, Wang2022tvt,liux2020tsp}.

\subsection{ISAC Signal Model}
Let $\mathbf{W}_{k}=[\mathbf{W}^c_{k},\mathbf{W}^r_{k}]\in\mathbb{C}^{N_t \times N_t}$ denote the beamforming matrix of ISAC device $k$, where $\mathbf{W}^c_{k}\in\mathbb{C}^{N_t \times d}$ and   $\mathbf{W}^r_{k}\in\mathbb{C}^{N_t \times (N_t-d)}$ represent the beamforming matrices for the communication symbols and radar waveforms, respectively. Here, $d$ denotes the number of data streams. The ISAC signal transmitted by ISAC device $k$ at time index $t$ is expressed as
\begin{equation}
\mathbf{x}_{k}(t) =\mathbf{W}_{k}[\mathbf{s}^c_{k}(t),\mathbf{s}^r_{k}(t)]^H= \mathbf{W}^c_{k}\mathbf{s}^c_{k}(t)+\mathbf{W}^r_{k}\mathbf{s}^r_{k}(t),
\end{equation}
where $\mathbf{s}^c_{k}(t)\in\mathbb{C}^{d \times 1}$ denotes device $k$'s communication data to be offloaded to the MEC server and $\mathbf{s}^r_{k}(t)\in\mathbb{C}^{(N_t-d) \times 1}$ denotes $(N_t-d)$ independently and pseudo-randomly generated radar waveform. According to \cite{liux2020tsp},  the radar waveforms are uncorrelated with the communication symbols, i.e., $\mathbb{E} \left\{ \mathbf{s}^r_k(t) (\mathbf{s}^{c}_k(t))^H\right\} = \mathbf{0}_{(N_t-d)\times d}$.

 The signal received by MEC server is expressed as
\begin{align}
\label{yk} 
\mathbf{y}(t)&=\mathbf{H}_{k}\mathbf{x}_{k}(t)+\sum_{i=1,i\neq k}^K\mathbf{H}_{i}\mathbf{x}_{i}(t) +\mathbf{n}_b(t),\notag\\
&=\mathbf{H}_{k}\mathbf{W}^c_{k}\mathbf{s}^c_{k}(t)+\sum_{i=1,i\neq k}^K\mathbf{H}_{i}\mathbf{W}^c_{k}\mathbf{s}^c_{i}(t)+\notag\\&~~~~~~\sum_{i=1}^K\mathbf{H}_{i}\mathbf{W}^r_{k}\mathbf{s}^r_{i}(t)+\mathbf{n}_b(t),
\end{align}
where $\mathbf{H}_k\in\mathbb{C}^{M \times N_t}$ denotes the channel matrix from ISAC device $k$ to MEC server and $\mathbf{n}_b(t)\sim \mathcal{CN}(0, {\sigma_b}^2\mathbf{I}_M )$ denotes the additive white Gaussian noise (AWGN) at the BS. We \textcolor{black}{consider that} all the channels are block-faded and  quasi-static; \textcolor{black}{hence, perfect channel state information (CSI) can be obtained \cite{Liu2018tsp}}. The maximum achievable rate from ISAC device $k$ to BS is given by
\begin{equation}
\label{r1}
R_{k} = B\log\det(\mathbf{I}_M+\mathbf{H}_{k}\mathbf{W}^c_{k}(\mathbf{W}^c_{k})^H\mathbf{H}^H_{k}\mathbf{D}^{-1}_{k} ).
\end{equation}

Here, $B$ denotes the signal bandwidth and $\mathbf{D}_{k}$ is given by
\begin{align}
\mathbf{D}_{lk}= &{\sum_{i=1,i\neq k}^K\mathbf{H}_{i}\mathbf{W}^c_{i}(\mathbf{W}^c_{i})^H\mathbf{H}^H_{i}}+\notag\\&~~\quad\quad{\sum_{i=1}^K\mathbf{H}_{i}\mathbf{W}^r_{i}(\mathbf{W}^r_{i})^H\mathbf{H}^H_{i}}+{\sigma_b}^2\mathbf{I}_M.
\end{align}

For the sensing perspective, we consider using the transmit beampattern as a key performance metric, which has been widely adopted in MIMO radar signal design. By properly designing the transmit beampattern, the system's sensing performance, including detection, estimation, and recognition, can be effectively enhanced. According to \cite{liux2020tsp}, the transmit beampattern is expressed as
\begin{equation}
\begin{aligned}\label{patt}
P_{b}(\theta) &=\mathbf{a}^H({\theta})\mathbf{R}_k\mathbf{a}({\theta}),
\end{aligned}
\end{equation}
where $\mathbf{a}(\theta) = [1, {\rm e}^{j2\pi \delta \sin(\theta)}, \ldots, {\rm e}^{j2\pi (N -1) \delta \sin(\theta)}]^T \in \mathbb{C}^{N_t \times 1}$ is the steering vector of transmit antennas, with $\delta$ denoting the normalized interval between adjacent antennas. Specifically, $\mathbf{R}_k$ denotes 
\begin{equation} \small \label{cov}
\mathbf{R}_k=\mathbb{E} \left\{ \mathbf{x}_k(t) \mathbf{x}^H_k(t)\right\} = \mathbf{W}_k\mathbf{W}^H_k=\mathbf{W}^c_k(\mathbf{W}^c_k)^H+\mathbf{W}^r_k(\mathbf{W}^r_k)^H. 
\end{equation}
 
From (\ref{patt}), we observe that realizing a desired transmit beampattern can be achieved by setting constraints for the covariance of the transmit beamforming matrix. Depending on specific sensing requirements, a desired beamforming covariance matrix can be pre-generated, denoted as $\hat{\mathbf{R}}^\text{pre}_{k}$\footnote{ \textcolor{black}{The desired  beamforming covariance $\hat{\mathbf{R}}^\text{pre}_{k}$ can be obtained by solving the constrained least-squares problem (19) in \cite{beam}, where  the desired transmit beampattern $\phi(\theta)$ is expressed as
	\begin{align} 
	\begin{split}
	\tilde \phi(\theta)= \left \{
	\begin{array}{ll}
	1,                    & |\theta-\theta_j|<\frac{\Delta \theta}{2},\\
	0,                    & \mathrm{otherwise},
	\end{array}
	\right.
	\end{split}
	\end{align}
	where $\theta_j, j = 1,..., J$ are the directions of the targets of interest and  $|\theta-\theta_j|<\frac{\Delta \theta}{2}$  defines the main-beam region of interest.}}. To attain high-quality sensing results, we impose constraints on the ISAC beamforming covariance matrices to match the predefined $\hat{\mathbf{R}}^\text{pre}_{k}$:
\begin{equation} \label{match}
 \mathbf{R}_k=\hat{\mathbf{R}}^\text{pre}_{k}, ~~\forall k. 
\end{equation}

\subsection{Computation Model}

\begin{figure*}[!t]
	\centering
	\includegraphics[width=5.7in, height=2.0in]{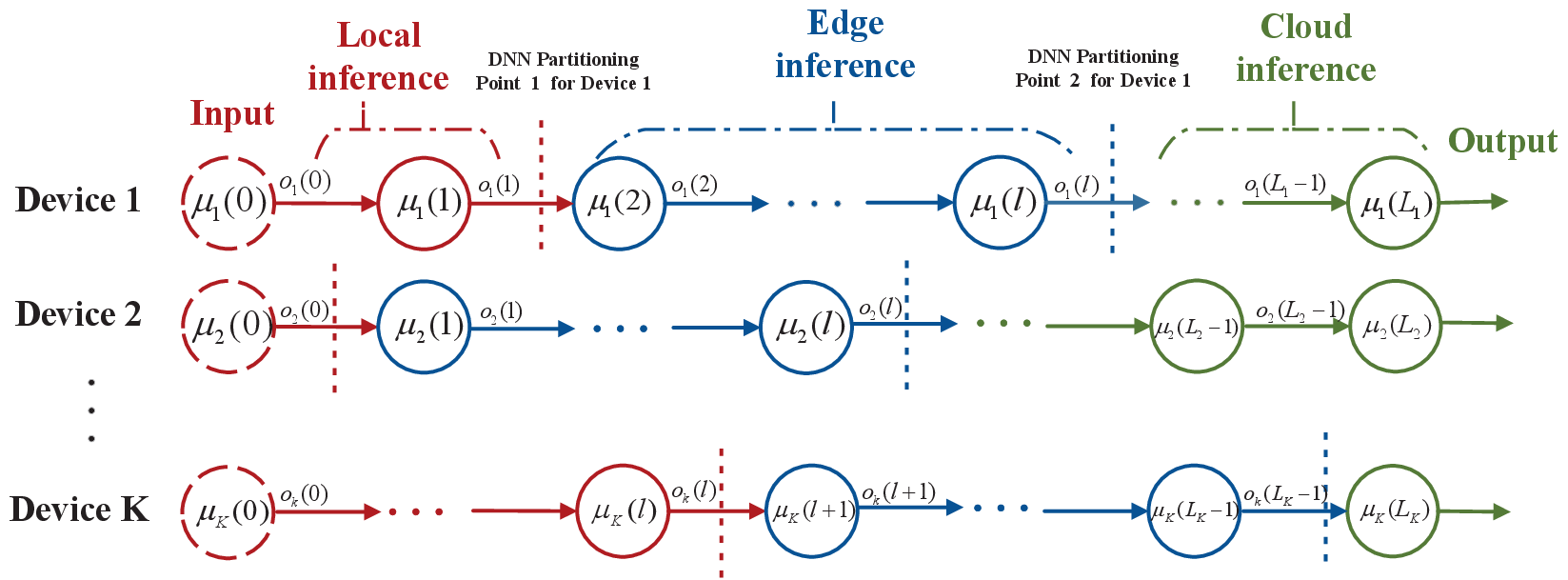}
	\caption{Three-Tier Collaborative Inference Framework.}
	\label{fig:frame}
\end{figure*}

In the computational process, we adopt a three-tier collaborative inference framework, as illustrated in Fig. {\ref{fig:frame}}, where a resource-constrained ISAC device executes a target classification task based on a pre-trained DNN with the assistance of an MEC server and a cloud server. Each entity executes a portion of the DNN to alleviate the computational burden on the ISAC device. Moreover, we use floating-point operations (FLOPs) to measure the computational resources required by each layer of the DNN model. For clarity, we denote the set of pre-trained DNN layers of ISAC device $k$ as $\mathcal{L}_k=\{1,\ldots,L_k\}$,  where the FLOPs required in the $l$-th layer are represented as $\mu_k(l)$, determined by the layer type $\zeta(l)$. The output data size (in bits) of the $l$-th layer is denoted as $o_k(l)$, which is fixed and dependent on  $\zeta(l)$. The computational load of DNN inference is primarily attributed to convolutional (CV) layers, fully connected (FC) layers, and max-pooling (MP) layers. The FLOPs $\mu_k(l)$ of each layer type is given by \cite{dnnmec1}
\begin{align}
\begin{split}
 \mu_k(l)= \left \{
\begin{array}{ll}
(2c(l-1)d(l)^2-1)a(l) b(l)c(l),                  & \zeta(l)=\text{CV},\\
a(l) b(l)c(l)d(l)^2,                    & \zeta(l)=\text{MP},\\
(2e(l-1)-1)e(l), & \zeta(l)=\text{FC},
\end{array}
\right.
\end{split}
\end{align}
where $a(l)$, $b(l)$, $c(l)$, $d(l)$, and $e(l)$ denote the height,  width, channels of the output feature map, the filter size, and the number of neurons at the $l$-th layer, respectively. To characterize the DNN partitioning strategy, we define the binary DNN partitioning decision variables as $\beta_{kl}\in\{0,1\}$, where $1\leq k\leq K, 0\leq l\leq L_k$. If layer $l$ in device $k$'s DNN model is selected as a partitioning point, then $\beta_{kl}=1$; otherwise, $\beta_{kl}=0$.  As illustrated in Fig. {\ref{fig:frame}}, the layer with $l=0$ represents a virtual input layer with $\mu_k(l)=0$. Since each device's DNN has at most two partitioning points to support cloud-edge-device collaborative inference, the following constraint holds
\begin{equation}
\begin{aligned}
1 \leq \sum_{l=1}^{L_k}\beta_{kl}\leq 2, \quad \forall k\in \mathcal{K}.
\end{aligned}
\end{equation}

By flexibly selecting partitioning points, the DNN model can be collaboratively performed by the ISAC devices, MEC server, and cloud server. The computational workloads (in FLOPs) required by ISAC device $k$, the MEC server, and the cloud server, denoted as $S^L_k$, $S^M_k$, and $S^C_k$, respectively, are given by
\begin{equation}\label{load}
\begin{aligned}
& S^L_k = \sum_{l = 0}^{l_1}\mu_k(l),~~S^M_k = \sum_{l = l_1+1}^{l_2}\mu_k(l), \\
& S^C_k = \sum_{l = 0}^{L_k}\mu_k(l)-S^L_k-S^M_k, \\&l_1, l_2 = \{ \hat{l} \mid \beta_{k\hat{l}} = 1 \},~~ l_1 \leq l_2,
\end{aligned}
\end{equation}
where $l_1$ and $l_2$ denote the first and second DNN partition layer indices corresponding to $\beta_{k\hat{l}}=1$, respectively.  Furthermore, if $l_1=l_2$,  we set $S^M_k=0$. 
\subsection{Cost Model}
Given a specific DNN partitioning strategy, the inference latency for executing the entire DNN model on device $k$ consists of five main components, as detailed below.
\subsubsection{Local Execution Latency}
Let $f^L_k$ denote the computation frequency (CPU cycles/s) of ISAC device $k$, and $\alpha^L_k$ represent its FLOPs per CPU cycle at device $k$. The inference latency for the portion  of the DNN executed on device $k$ can be expressed as
\begin{equation}\label{t1}
T_{1,k}^C=\frac{{S}^L_k}{\alpha^L_kf^L_k}.
\end{equation}
\subsubsection{ISAC Device-MEC Server Offloading Latency}
According to (\ref{r1}) and (\ref{load}), the offloading latency from ISAC device $k$ to the MEC server can be expressed as
\begin{equation}\label{t2}
T_{1,k}^O=\frac{{o}_k(l_1)}{R_k}.
\end{equation}
\subsubsection{MEC Server Execution Latency}
Let $f^M_k$ denote the computational resources (CPU cycles/s) allocated by the MEC server to device $k$, and $\alpha^M_k$ denote the FLOPs per CPU cycle at the MEC server. Then, the inference latency for the portion of the DNN executed on the MEC server is expressed as
\begin{equation}\label{t3}
T_{2,k}^C=\frac{{S}^M_k}{\alpha^M_kf^M_k}.
\end{equation}

Since the computational resources of the MEC server are typically limited, we have 
\begin{equation}
\sum_{k=1}^K{f}_k^M \leq {F}_M,
\end{equation}
where $F_M$ is the computation capacity (CPU cycles/s) of MEC server.
\subsubsection{Transmission Latency From MEC Server to Cloud Server }
We assume that the  MEC server is connected to the cloud server via optical fiber, with a backhaul rate of $r_b$. Then, the transmission delay from the MEC server to the cloud server is given by
\begin{equation}\label{t4}
T_{2,k}^O=\frac{{o}_k(l_2)}{r_b}.
\end{equation}
\subsubsection{Cloud Server Execution Latency}
Since the cloud server has sufficient computational resources, we assume that the computational resources allocated to each device are fixed as $f^C$. Let $\alpha^C_k$ denote the FLOPs per CPU cycle at the cloud server.  The inference latency for the portion of the DNN executed on the cloud server is given by
\begin{equation}\label{t5}
T_{3,k}^C=\frac{{S}^C_k}{\alpha^C_kf^C}.
\end{equation}

It is worth noting that before being fed into the DNN model, the sensing data undergoes preprocessing steps such as sampling, filtering, and STFT to be converted into a spectrogram. Since these preprocessing operations, which have relatively low computational complexity, can be efficiently executed on an hardware module, their processing latency is negligible compared to the DNN computation and data transmission~\cite{dnnmec3, dnniscc3}. Furthermore, as the size of the classification results is much smaller than that of the input spectrogram, the feedback latency is similarly negligible, as noted in~\cite{ISCC2,ISCC5}. Consequently, the  total inference latency is expressed as
\begin{equation}\label{t6}
\begin{aligned}
T^{I}_k&=T_{1,k}^C+T_{1,k}^O+T_{2,k}^C+T_{2,k}^O+T_{3,k}^C,\\
&=\frac{{S}^L_k}{\alpha^L_kf^L_k}+\frac{{o}_k(l_1)}{R_k}+\frac{{S}^M_k}{\alpha^M_kf^M_k}+\frac{{o}_k(l_2)}{r_b}+\frac{{S}^c_k}{\alpha^C_kf^C}.
\end{aligned}
\end{equation}

Note that the DNN computation typically incurs high energy consumption while  devices have limited energy; hence, their computational energy usage must be carefully considered\footnote{The MEC and cloud servers are typically powered by stable wired connections, and \textcolor{black}{hence they typically} have a sufficient energy supply.}. The power consumption model of computation is given by \cite{ISCC2}
\begin{equation}
 p_k=\kappa(f^L_k)^3,
 \end{equation} 
 where $\kappa$ is a coefficient determined by the hardware architecture. Consequently, the computational energy consumption of device $k$ can be expressed as
\begin{equation}\label{e1}
E_k=p_kT^C_{1,k}=\frac{\kappa{S}^L_k(f^L_k)^2}{\alpha_k^L}.
\end{equation}

\subsection{Problem Formulation}
Based on the above modeling,  the ISAC beamforming matrix $\{\mathbf{W}_k\}_{\forall k}$ influences not only the sensing performance but also the offloading latency.  The allocation of computational resources, $\bm{f}^M=\{f_k^M\}_{\forall k}$ and $\bm{f}^L=\{f_k^L\}_{\forall k}$, affects both MEC server and local execution latency. As shown in (\ref{load}) and (\ref{t6}), the DNN partitioning decision variables $\bm{\beta}=\{\beta_{kl}\}_{\forall k,l}$ impact the distribution of computational load across the cloud, MEC servers, and devices, as well as the amount of data transferred among them. Consequently, $\bm{\beta}$, $\{\mathbf{W}_k\}_{\forall k}$,  $\bm{f}^M$, and  $\bm{f}^L$  jointly determine the total inference latency of the DNN model. In this paper, we aim to minimize the overall inference latency of all devices' DNN-based sensing tasks, subject to sensing beampattern matching constraints (\ref{match}). To this end, we jointly optimize the DNN partitioning decision variables $\bm{\beta}$, the beamforming matrix $\{\mathbf{W}_k\}_{\forall k}$, and the allocation of computational resources between the MEC server and devices, $\bm{f}^M$ and $\bm{f}^L$. The optimization problem is formulated  as follows:
\begin{align} 
\label{prob29}
\min_{\substack{{\bm{\beta}, \mathbf{W}_k},\\ \bm{f}^M,\bm{f}^L}} & ~\sum_{k=1}^{K}~T_{k}^{I}\\
s.t.~
&1 \leq \sum_{l=1}^{L_k}\beta_{kl}\leq 2, \quad \forall k\in \mathcal{K}, \tag{\ref{prob29}a}\\
& \beta_{kl} \in \left\{ {0,1} \right\},~\forall k\in \mathcal{K}, l\in \mathcal{L}_k, \tag{\ref{prob29}b}\\
&\sum_{k=1}^K{f}_k^M \leq {F}_M, \tag{\ref{prob29}c}\\
& 0\leq{f}_k^L\leq{F}_k, ~~~\forall k\in \mathcal{K}, \tag{\ref{prob29}d}\\
&E_k \leq E_{th}, ~~\forall k\in \mathcal{K},\tag{\ref{prob29}e}\\
&\mathbf{W}_k\mathbf{W}^H_k=\hat{\mathbf{R}}^\text{pre}_{k}, ~~\forall k\in \mathcal{K},\tag{\ref{prob29}f}
\end{align}
where $F_k$ and $E_{th}$ denote the computational capacity and the computational energy budget of ISAC device $k$, respectively. Due to the intricate coupling between variables $\bm{\beta}$, $\{\mathbf{W}_k\}_{\forall k}$, $\bm{f}^M$ and $\bm{f}^L$, the presence of fractional expressions in the objective function, and the quadratic equality constraints in (\ref{prob29}f), problem (\ref{prob29}) is highly non-convex and difficult to solve. The inclusion of integer variables and index-dependent terms in (\ref{load}) further complicates the optimization. In particular, when $\{\mathbf{W}_k\}_{\forall k}$,  $\bm{f}^M$, and  $\bm{f}^L$ are fixed,  the generalized form of (\ref{prob29}) reduces to a multi-choice knapsack problem, which is known to be NP-hard. This leads to a high-dimensional decision space for optimizing the DNN partitioning strategy. Specifically, determining the optimal $\bm{\beta}$ requires exhaustively searching through all $(\frac{L_k(L_k+1)}{2})^K$ possible cases, which becomes computationally infeasible as $L_k$ and $K$ increase. Therefore, solving the original mixed-integer nonlinear programming (MINLP) problem in (\ref{prob29}) poses significant computational challenges.

\section{Proposed Two-Layer Optimization Algorithm}
In this section, we propose an efficient two-layer optimization algorithm to solve the MINLP problem (\ref{prob29}). Specifically, we first leverage the hierarchical structure of the problem to  decouple it into 
\begin{itemize}
	\item Inner-layer optimization problem: Optimizes the beamforming matrices  $\{\mathbf{W}_k\}_{\forall k}$, MEC server computational resource allocation $\bm{f}^M$,  and local computational resource allocation  $\bm{f}^L$.
	\item Outer-layer optimization problem: Optimizes the DNN partitioning decision variables $\bm{\beta}$.
\end{itemize}

For the inner-layer optimization problem, we further decompose it into three independent subproblems and derive closed-form solutions for $\bm{f}^M$,  $\bm{f}^L$ and $\{\mathbf{W}_k\}_{\forall k}$ using the KKT conditions, MM framework, and WMMSE-based block coordinate descent (BCD) method, respectively. To address the outer-layer optimization problem, we propose a CE-based algorithm that iteratively learns the optimal probability distribution of the DNN partitioning decision variables. Specifically, this is accomplished by randomly sampling $\bm{\beta}$, solving the corresponding inner-layer problem for each sample, and updating the distribution based on the resulting objective values, thereby progressively guiding the search toward the optimal solution.

\subsection{Problem Decomposition}
By leveraging the layered structure of different types of variables, problem (\ref{prob29}) can be vertically decomposed into two subproblems. Given the DNN partitioning decision variables  $\bm{\beta}$, the inner-layer optimization problem with respect to the continuous variables $\bm{f}^M$,  $\bm{f}^L$, and $\{\mathbf{W}_k\}_{\forall k}$ can be formulated as follows:
\begin{align} 
\label{prob23}
\min_{\substack{{  \bm{f}^M,\bm{f}^L}\\\mathbf{W}_k}} &  ~~\sum_{k=1}^{K}~T_{k}^{I}\\
s.t.~&~~ \text{(\ref{prob29}\text{c})},\text{(\ref{prob29}\text{d})},\text{(\ref{prob29}\text{e})},\text{(\ref{prob29}\text{f})}. \notag
\end{align}

After solving the inner-layer optimization problem, the outer-layer optimization problem with respect to the discrete variable $\bm{\beta}$ can be formulated as
\begin{align} 
\label{prob4}
\min_{\bm{\beta} } &  ~~\sum_{k=1}^{K}~T_{k}^{I}\\
s.t.& ~\text{(\ref{prob29}\text{a})},\text{(\ref{prob29}\text{b})}. \notag
\end{align}

\subsection{Solution of Inner-layer Optimization Problem}
For the inner-layer problem (\ref{prob23}), we observe that the objective function terms and the constraints associated with the continuous variables $\bm{f}^M$,  $\bm{f}^L$, and $\{\mathbf{W}_k\}_{\forall k}$ are independent to each other. Therefore, the problem can be decomposed into three separate subproblems, with each solved individually.

\subsubsection{Optimization of MEC Computational Resource Allocation}
The subproblem with respect to the MEC computational resource allocation $\bm{f}^M$ is  formulated as
\begin{align} 
\label{prob5}
\min_{\bm{f}^M } &  ~~\sum_{k=1}^{K}~\frac{{S}^M_k}{\alpha^M_kf^M_k}\\
s.t.~& ~~\text{(\ref{prob29}\text{c})}, \notag
\end{align}
which is a standard convex optimization problem that can be directly solved using well-established convex optimization toolboxes such as CVX. To reduce computational complexity, we derive its optimal closed-form solution using KKT condition.  Specifically, the  Lagrangian function of (\ref{prob5}) is formulated as
\begin{equation}
\begin{aligned}
\label{lg1}
&\mathcal{L}\left( \bm{f}^M, \eta \right)= \sum_{k=1}^{K}~\frac{{S}^M_k}{\alpha^M_kf^M_k}+\eta\left(\sum_{k=1}^K{f}_k^M-F_M\right)
\end{aligned}
\end{equation}
where $\eta$ is the a Lagrange multiplier associated with the MEC server's computational capacity constraint. Based on the first-order optimality in  KKT conditions \cite{2004Convex}, the optimal closed-form solution of ${f}^M_{k}$  is derived as follows:
\begin{equation}  \label{fm}
{f}^{M*}_{k}=\sqrt{\frac{{S}^M_k}{\alpha^M_k\eta^*}}, \quad \forall k\in \mathcal{K},
\end{equation}
where $\eta^*$ is chosen such that the complementary slackness condition for the computational capacity constraint is satisfied, i.e., $\eta^*\left(\sum_{k=1}^K{f}_k^{M*}-F_M\right)=0$. Note that when $\eta^ *=0$, the optimal computational frequency in (\ref{fm}) tends to infinity, which is impractical. Thus, we have $\sum_{k=1}^K{f}_k^{M*}-F_M=0$ and the optimal $\eta^*$ is given by
\begin{equation}   \label{fm2}
\eta*=\left(\sum_{k=1}^K\frac{\sqrt{{S}^M_k}}{\sqrt{\alpha^M_k}F_M}\right)^2.
\end{equation}

Finally, substituting (\ref{fm2}) into (\ref{fm}),  the optimal ${f}^M_{k}$ is 
\begin{equation}  \label{fm3}
{f}^{M*}_{k}=\frac{\sqrt{{S}^M_k}}{\sum_{i=1}^K\sqrt{\frac{{\alpha^M_k}{S}^M_i}{\alpha^M_i}}}F_M.
\end{equation}

\subsubsection{Optimization of Local Computational Resource Allocation}
The subproblem for local computational resource allocation $\bm{f}^L$ is given by
\begin{align} 
\label{prob6}
\min_{\bm{f}^L } &  ~~\sum_{k=1}^{K}~\frac{{S}^L_k}{\alpha^L_kf^L_k}\\
s.t.& ~~\text{(\ref{prob29}\text{d})},\text{(\ref{prob29}\text{e})}, \notag
\end{align}
which can be decomposed into $K$ independent linear programming problems, given by
\begin{align} 
\label{prob7}
{\mathcal{P}}_k^{\mathrm{sub1}}(f_k^L):\max_{f^L_k } &  ~~\frac{\alpha^L_kf^L_k}{{S}^L_k}\\
s.t.~& 0\leq{f}_k^L\leq{F}_k, \tag{\ref{prob7}a}\\
&\frac{\kappa{S}^L_k(f^L_k)^2}{\alpha_k^L}\leq E_{th}. \tag{\ref{prob7}b}
\end{align}

Exploiting the linearity of (\ref{prob7}), the optimal $f_k^L$ is derived as
\begin{equation}  \label{fm4}
{f}^{L*}_{k} = \min\left\{F_k, \sqrt{\frac{E_{th}\alpha_k^L}{\kappa{S}^L_k}}\right\}.
\end{equation}

\subsubsection{Optimization of Beamforming Matrices}
The subproblem for the ISAC beamforming matrix is formulated as
\begin{align} 
\label{prob8}
\min_{\mathbf{W}_k} &  ~~\sum_{k=1}^{K}~\frac{{o}_k(l_1)}{R_k}\\
s.t.~& ~~\text{(\ref{prob29}\text{f})}, \notag
\end{align}
which is  a sum-of-ratios minimization problem and cannot be directly solved using classical fractional programming methods. To address this, we adopt the MM  framework, which first constructs a surrogate function that locally approximates the objective function, ensuring that their difference is minimized at the current point. Then, by iteratively minimizing the surrogate function, the original problem can be efficiently solved \cite{MM}. Specifically, we first reformulate the objective of (\ref{prob8})  as
\begin{align} 
\label{prob8_2}
\max_{\mathbf{W}_k} &  ~~\frac{1}{\sum_{k=1}^{K}~\frac{{o}_k(l_1)}{R_k}}.
\end{align}

According to Jensen's inequality \cite{MM}, the lower-bound
surrogate function is given by
\begin{equation}  \label{MM}
\frac{1}{\sum_{k=1}^{K}\frac{{o}_k(l_1)}{R_k}}\geq \frac{\sum_{k=1}^{K}\frac{{o}_k(l_1)}{(R^{it}_k)^2}R_k}{\left(\sum_{k=1}^{K}\frac{{o}_k(l_1)}{R^{it}_k}\right)^2}=\sum_{k=1}^{K}z_k^{it}R_k,
\end{equation}
where $R^{it}_k$ represents the value of $R_k$ at ${it}$-th MM iteration and $\{z^{it}_k\}_{\forall k}$ are introduced auxiliary variables that remain fixed in the $(it+1)$-th iteration. Subsequently, the problem to be solved at each MM iteration is formulated as
\begin{align} 
\label{prob10}
\max_{\mathbf{W}_k} &  ~~\sum_{k=1}^{K}~z^{it}_{k}{R_{k}}\\
s.t.~& ~~\text{(\ref{prob29}\text{f})}. \notag
\end{align}

We observe that problem (\ref{prob10}) remains non-convex. To address this, we propose a BCD algorithm based on the WMMSE approach \cite{WMMSE}. Specifically, considering that the BS applies a linear receiver $\mathbf{U}_k \in \mathbb{C}^{d \times M}$  for ISAC device $k$’s signal detection, the mean square error (MSE) matrix is given by
\begin{equation}
\begin{aligned}
\label{mse}
\mathbf{E}_k&=\mathbb{E}\left[\left(\mathbf{U}_k \mathbf{y}_k(t)-\mathbf{s}^c_k(t)\right)\left(\mathbf{U}_k \mathbf{y}_k(t)-\mathbf{s}^c_k(t)\right)^{\mathrm{H}}\right] \\
&=\mathbf{I}_d-2\text{Re}\{\mathbf{U}_k\mathbf{H}_k\mathbf{W}^c_k\}+{\sum_{i=1}^K\mathbf{U}_k\mathbf{H}_{i}\mathbf{W}^c_{i}(\mathbf{W}^c_{i})^H\mathbf{U}_k^H}\\
&\quad~~+{\sum_{i=1}^K\mathbf{U}_k\mathbf{H}_{i}\mathbf{W}^r_{i}(\mathbf{W}^r_{i})^H\mathbf{H}^H_{i}\mathbf{U}_k^H}+{\sigma_b}^2\mathbf{U}_k\mathbf{U}_k^H,\\
&=\mathbf{I}_d-2\text{Re}\{\mathbf{U}_k\mathbf{H}_k\mathbf{W}^c_k\}+\\
&\quad~~+{\sum_{i=1}^K\mathbf{U}_k\mathbf{H}_{i}\hat{\mathbf{R}}^\text{pre}_{k}\mathbf{H}^H_{i}\mathbf{U}_k^H}+{\sigma_b}^2\mathbf{U}_k\mathbf{U}_k^H.
\end{aligned}
\end{equation}

According to \cite{WMMSE}, the weighted sum-rate maximization problem (\ref{prob10}) can be equivalently transformed into the following matrix-weighted sum-MSE minimization problem 
\begin{align} 
\label{prob11}
\min_{\mathbf{U}_k,\mathbf{G}_k,\mathbf{W}_k} &  ~~\sum_{k=1}^{K}z^{it}_{k}(\text{Tr}(\mathbf{G}_k\mathbf{E}_k)-\log\det(\mathbf{G}_k))\\
s.t.~& ~~\text{(\ref{prob29}\text{f})}, \notag
\end{align}
where $\mathbf{G}_k\succeq0, \forall k\in \mathcal{K}$ are the auxiliary  variables. Next, we alternately optimize $\mathbf{U}_k$, $\mathbf{G}_k$, and $\mathbf{W}_k$ based on the BCD technique. For the optimization of $\mathbf{U}_k$, given $\mathbf{G}_k$ and $\mathbf{W}_k$, the optimal receiver is the well-known MMSE receiver, which is given by
   \begin{equation}
	\label{MMSE1}
	\mathbf{U}^*_k=(\mathbf{W}^c_{k})^H\mathbf{H}^H_{k}\left(\sum_{i=1}^K\mathbf{H}_{i}\hat{\mathbf{R}}^\textnormal{pre}_{k}\mathbf{H}^H_{i}+{\sigma_b}^2\mathbf{I}_M\right)^{-1}.
	\end{equation}

  For the optimization of $\mathbf{G}_k$, we substitute (\ref{MMSE1}) into (\ref{mse}), and the MSE matrix can be reformulated as
\begin{equation}
\mathbf{E}^*_k = \mathbf{I}_d-(\mathbf{W}^c_{k})^H\mathbf{H}^H_{k}\left(\sum_{i=1}^K\mathbf{H}_{i}\hat{\mathbf{R}}^\text{pre}_{k}\mathbf{H}^H_{i}+{\sigma_b}^2\mathbf{I}_M\right)^{-1}\mathbf{H}_{k}\mathbf{W}^c_{k}.
\end{equation}

Given fixed $\mathbf{U}_k$ and $\mathbf{W}_k$, the problem (\ref{prob11}) is an  unconstrained convex problem. The optimal $\mathbf{G}_k$ can be obtained using the  first-order optimality in  KKT conditions  as follows:
\begin{equation} 	\label{MMSE3}
\mathbf{G}^*_k=(\mathbf{E}^*_k)^{-1}.
\end{equation}

 Next, given $\mathbf{U}_k$ and $\mathbf{G}_k$, we substitute (\ref{mse}) into (\ref{prob11}) and reformulate it as
 \begin{align} 
 \label{prob12}
 \max_{\mathbf{W}_k} &  ~~\sum_{k=1}^{K}z^{it}_{k}\text{Re}\{\text{Tr}\left(\mathbf{G}_k\mathbf{U}_k\mathbf{H}_k\mathbf{W}^c_k\right) \} \\
 s.t.~& ~~\text{(\ref{prob29}\text{f})}. \notag
 \end{align}
 
Unlike the convex beamforming optimization subproblem in \cite{WMMSE}, which can be directly solved via KKT conditions or interior-point methods \cite{2004Convex}, the presence of non-convex quadratic equality constraints renders the problem non-convex and difficult to solve. To address this, we reformulate it as an OPP and derive its optimal closed-form solution.  Specifically, we first apply Cholesky decomposition $\hat{\mathbf{R}}^\text{pre}_{k}=\mathbf{Q}_k\mathbf{Q}_k^H$ and define the auxiliary variables $\widehat{\mathbf{W}}_k=\mathbf{Q}^{-1}_k{\mathbf{W}}_k$ and $\widehat{\mathbf{H}}_k={\mathbf{H}}_k\mathbf{Q}_k$. As a result,  the problem  (\ref{prob12}) can be decomposed into $K$ independent subproblems, formulated as follows:
  \begin{align} 
  \label{prob13}
 {\mathcal{P}}_k^{\mathrm{sub2}}(\widehat{\mathbf{W}}_k):\max_{\widehat{\mathbf{W}}_k} &  ~~\text{Re}\left\{\text{Tr}\left(\mathbf{T}^H_k\widehat{\mathbf{W}^c_k}\right) \right\} \\
 s.t.~&~~\widehat{\mathbf{W}}_k\widehat{\mathbf{W}}^H_k=\mathbf{I}_{N_t},\tag{\ref{prob13}a}
\end{align}
 where $\mathbf{\mathbf{T}_k}=\widehat{\mathbf{H}}^H_k\mathbf{U}^H_k\mathbf{G}^H_k$. Since the objective function of (\ref{prob13}) depends only on $\widehat{\mathbf{W}^c_k}$, it can be rewritten as
   \begin{align} 
 \label{prob14}
 {\mathcal{P}}_k^{\mathrm{sub2}}(\widehat{\mathbf{W}}^c_k):\max_{\widehat{\mathbf{W}}^c_k} &  ~~\text{Re}\left\{\text{Tr}\left(\mathbf{T}^H_k\widehat{\mathbf{W}^c_k}\right) \right\} \\
 s.t.~&~~(\widehat{\mathbf{W}}^c_k)^H\widehat{\mathbf{W}}^c_k=\mathbf{I}_{d}.\tag{\ref{prob14}a}
 \end{align}
 
 According to constraint (\ref{prob13}a), we have $\widehat{\mathbf{W}}^r_k(\widehat{\mathbf{W}}^r_k)^H=\mathbf{I}_{N_t-d}$ and $\widehat{\mathbf{W}}^r_k(\widehat{\mathbf{W}}^c_k)^H=\bm{0}$.  Once the optimal $\widehat{\mathbf{W}}^{c*}_k$ is obtained,   the corresponding $\widehat{\mathbf{W}}^{r*}_k$ can be directly derived from these orthogonal constraints. To address problem (\ref{prob14}),  we present the following lemma.
\begin{lemma}
	Problem (\ref{prob14}) can be equivalently transformed into the following OPP
	   \begin{align} 
	 \label{prob15}
	{\mathcal{P}}_k^{\mathrm{sub3}}(\widehat{\mathbf{W}}^c_k):\min_{\widehat{\mathbf{W}}^c_k} &  ~~||\mathbf{T}_k-\widehat{\mathbf{W}}^c_k||_F^2 \\
	 s.t.~& ~~\textnormal{(\ref{prob14}\text{a})}. \notag
\end{align}
\end{lemma}
\begin{IEEEproof}
When $ \mathbf{U}_k$ and $ \mathbf{G}^H_k$ are fixed, $\mathbf{\mathbf{T}_k}=\widehat{\mathbf{H}}^H_k\mathbf{U}^H_k\mathbf{G}^H_k$ is constant. According to the constraint (\ref{prob14}a), we have $\text{Tr}\left((\widehat{\mathbf{W}}^c_k)^H\widehat{\mathbf{W}}^c_k\right)=d$. 
By additionally introducing two constant terms $\text{Tr}\left(\widehat{\mathbf{T}}_k^H\widehat{\mathbf{T}}_k\right)$ and $\text{Tr}\left((\widehat{\mathbf{W}}^c_k)^H\widehat{\mathbf{W}}^c_k\right)$, the objective function in (\ref{prob14}) can be equivalently transformed into
\begin{equation}
\begin{aligned} 
\label{prob18}
\max_{\widehat{\mathbf{W}}^c_k} &  ~\text{Re}\left\{\text{Tr}\left(\mathbf{T}^H_k\widehat{\mathbf{W}}^c_k\right) \right\} = 
\min_{\widehat{\mathbf{W}}^c_k}   -\text{Re}\left\{\text{Tr}\left(\mathbf{T}^H_k\widehat{\mathbf{W}}^c_k\right) \right\}  \\
=\min_{\widehat{\mathbf{W}}^c_k}  & ~\text{Tr}\left(\widehat{\mathbf{T}}_k^H\widehat{\mathbf{T}}_k-2\text{Re}\left\{\mathbf{T}^H_k\widehat{\mathbf{W}}^c_k\right\}+(\widehat{\mathbf{W}}^c_k)^H\widehat{\mathbf{W}}^c_k\right) \\ 
=\min_{\widehat{\mathbf{W}}^c_k} &  ~||\mathbf{T}_k-\widehat{\mathbf{W}}^c_k||_F^2. 
\end{aligned}
\end{equation}
\end{IEEEproof}

According to \cite{OOP}, the optimal closed-form solution to the OPP can be obtained via singular value decomposition (SVD). Specifically, by performing SVD on $\mathbf{T_k}$ as $\mathbf{T_k}=\mathbf{A} \mathbf{\Sigma}\mathbf{B}^H$, the optimal $\widehat{\mathbf{W}}^{c*}_k$ is given by
\begin{equation}\label{w1}
\widehat{\mathbf{W}}^{c*}_k=\mathbf{A}(1:d)\mathbf{B}^H,
\end{equation}
where $\mathbf{A}(1:d)$ represents the first $d$ left singular vectors corresponding to the top $d$ singular values of $\mathbf{\Sigma}$. Next, given the orthogonality constraints $\widehat{\mathbf{W}}^r_k(\widehat{\mathbf{W}}^r_k)^H=\mathbf{I}_{N_t-d}$ and $\widehat{\mathbf{W}}^r_k(\widehat{\mathbf{W}}^c_k)^H=\bm{0}$, the optimal $\widehat{\mathbf{W}}^{r*}_k$ can be obtained as follows:
\begin{equation}\label{w2}
\widehat{\mathbf{W}}^{r*}_k=\mathbf{A}(d+1:N_t),
\end{equation}
where $\mathbf{A}(d+1:N_t)$ represents the remaining $N_t-d$ left singular vectors. Finally, the optimal beamforming matrices are recovered as
${\mathbf{W}}^{c*}_k=\mathbf{Q}\widehat{\mathbf{W}}^{c*}_k$ and ${\mathbf{W}}^{r*}_k=\mathbf{Q}\widehat{\mathbf{W}}^{r*}_k$.

  \begin{algorithm}[t] 
	\caption{The Overall algorithm for solving inner-layer optimization  problem  (\ref{prob23})  }
	\begin{algorithmic}[1]
		\renewcommand{\algorithmicrequire}{\textbf{Input:}}
		\renewcommand{\algorithmicensure}{\textbf{Output:}}
		\Require The DNN partitioning decision variables $\bm{\beta}$, maximum iteration number $it_{max}$,  the convergence threshold $\epsilon>0$,
		\State {Obtain optimal MEC computational resource allocation $\bm{f}^M$ based on (\ref{fm3}).}
		\State {Obtain optimal local computational resource allocation $\bm{f}^L$ based on (\ref{fm4}).}
		\State {Initialize   $\widehat{\mathbf{W}}^*_k$,  $it=0$, $R^{it}_k$ and $z^{it}_k$.}
		\Repeat
		\Repeat
		\State {Update $\mathbf{U}^*_k$ based on (\ref{MMSE1}). }
		\State {Update  $\mathbf{G}^*_k$ based on (\ref{MMSE3}).}
		\State {Update  $\widehat{\mathbf{W}}^*_k$ based on (\ref{w1}) and (\ref{w2}), and obtain ${\mathbf{W}}^*_k=\mathbf{Q}\widehat{\mathbf{W}}^*_k$.}

		\Until {The objective value of (\ref{prob10}) converges.}
			\State {Update ${it} = {it} + 1$.}
			\State {Update  $R^{it}_k$ and $z^{it}_k$.}
			\Until {The objective value of (\ref{prob8}) converges or $it>it_{max}$.}
		\Ensure The MEC server computational resource allocation $\bm{f}^M$,  local computational resource allocation  $\bm{f}^L$, and the beamforming matrix  $\{\mathbf{W}_k\}_{\forall k}$.
	\end{algorithmic}
\end{algorithm}

\subsubsection{Overall Algorithm for the Inner-Layer Optimization  Problem}
To enhance clarity, we summarize the overall algorithm for solving the inner-layer optimization problem (\ref{prob23}) in Algorithm 1. Specifically, we first decompose problem (\ref{prob23}) into three independent subproblems. Then, we derive the optimal closed-form solutions for MEC computational resource allocation $\bm{f}^M$ and local computational resource allocation $\bm{f}^L$ by leveraging the KKT conditions and the linear structural properties of the optimization problem. Finally, we adopt the MM framework and propose a WMMSE-based BCD algorithm, incorporating an OPP transformation, to derive the optimal closed-form solution for the beamforming matrices $\mathbf{W}_k$.

 \subsection{Solution of Outer-layer Optimization Problem}
Next, we address the outer-layer optimization problem (\ref{prob4}), which involves determining the DNN partitioning decision variables $\bm{\beta}$. Due to the binary constraints in (\ref{prob29}\text{a}) and (\ref{prob29}\text{b}), this problem constitutes an integer program and cannot be solved using conventional convex optimization techniques. A typical strategy is to relax the binary variables $\beta_{kl} \in \{0,1\}$ to continuous values within $[0,1]$, transforming the problem into a convex form \cite{liu2024joint}. However, this relaxation is not well-suited to our setting for two main reasons. First, the relaxed solution is generally suboptimal and may lead to significant performance loss. Second, even with relaxation, the objective function and the index-dependent expressions in (\ref{load}) remain non-convex. While exhaustive search or BnB methods can theoretically yield the global optimum, their computational complexity becomes prohibitive as the number of decision variables grows. To overcome this challenge, we adopt a probabilistic model-based approach inspired by the CE method in machine learning (ML), which efficiently learns the optimal probability distribution of each DNN partitioning variable $\beta_{kl}$.
 \subsubsection{The Concept of CE} 
  Cross-entropy is widely used in probability theory to measure the difference between two probability distributions. Given two distributions $f(x)$ and $g(x)$, their cross-entropy is defined as 
\begin{equation} \label{CE1}
D(f \| g) = \mathbb{E}_f \left[ \ln \frac{f(x)}{g(x)} \right]=
\sum f(x) \ln f(x) 
{- \sum f(x) \ln g(x)}.
\end{equation}

In ML field, problems are typically modeled within a probabilistic distribution function learning framework, aiming to find the optimal input probability distribution that best matches the desired output using the training dataset. In this context, $f(x)$ represents the empirical distribution that characterizes the true distribution of the optimal solution, while  $g(x)$ denotes the theoretically tractable probability distribution model that we attempt to learn from the training dataset.

\subsubsection{The CE-based Algorithm} 
Building on the concept of CE, we reformulate the outer-layer optimization problem over $\beta_{kl}$ as a CE minimization problem. Noting that the first term in (\ref{CE1}) is constant with respect to the target distribution $g(x)$, the minimization can be equivalently expressed as:
\begin{align} 
\label{CE2}
\min_{g(\bm{\beta})} &  ~~-\sum f(\bm{\beta}) \ln g(\bm{\beta})\\
s.t.~& ~~\text{(\ref{prob29}\text{a})},\text{(\ref{prob29}\text{b})}. \notag
\end{align}

To solve this problem, we adopt the Monte Carlo simulation method, which mainly consists of  four steps:
\begin{itemize}
 \item Generating random samples of DNN partitioning decision variables based on the initialized probability function $g(\bm{\beta})$.
 \item Selecting high-performing ``elite samples" for probability learning and updating $g(\bm{\beta})$; 
 \item Iteratively generating new samples-based updated $g(\bm{\beta})$ and learning the optimal probability distribution until convergence; 
 \item Deriving the optimal DNN partitioning strategy that minimizes inference latency based on the well-learned $g(\bm{\beta})$.
\end{itemize}

In CE-based algorithms, the choice of a theoretically tractable probability distribution function $g( \bm{\beta})$ strongly depends on the type of optimization variables. Since the DNN partitioning decision variable is a binary integer variable, we model it using a Bernoulli distribution, i.e., $ \bm{\beta}\sim \text{Bernoulli}(\bm{\omega})$ \cite{CE}, where $\bm{\omega}=\{\omega_{kl}\}_{\forall k,l}$ is the parameter
vector denoting the probability that $\{\beta_{kl}=1\}_{\forall k,l}$.  Accordingly, the distribution of the optimization variable $ \bm{\beta$} is
\begin{equation} \label{CE3}
g(\bm{\beta}, \bm{\omega}) = \prod_{k=1}^{K}\prod_{l=1}^{L_k} \omega_{kl}^{\beta_{kl}} (1 - \omega_{kl})^{(1 - \beta_{kl})}.
\end{equation}

We begin by assuming that $ \beta_{kl}=1$ and $ \beta_{kl}=0$ are equally likely, and initialize the probability vector as $ \bm{\omega}=0.5\times\bm{1}$, where $\bm{1}$ denotes the all-ones vector. Based on the probability distribution in (\ref{CE3}), we generate $V$ feasible samples that satisfy the constraints (\ref{prob29}\text{a}) and (\ref{prob29}\text{b}) through stochastic sampling. For each sample, the corresponding inner-layer problem (\ref{prob23}) is solved to obtain the objective value of the original problem (\ref{prob29}).  These samples are then sorted in ascending order according to their objective values, and the top $V_{\text{elite}}$ samples are selected as elite candidates, denoted by $\bm{V}{\text{elite}} = {\bm{\beta}^{[1]}, \dots, \bm{\beta}^{[{V}{\text{elite}}]}}$, for learning the optimal distribution. Since the empirical distribution $f(\bm{\beta})$ of the optimal solution is unknown a priori, a widely adopted approach is to treat all elite samples as equally likely high-quality solutions, i.e., $f(\bm{\beta})=\frac{1}{{V}_{elite}}$ if $\bm{\beta}\in\bm{V}_{elite}$, and  $f(\bm{\beta})=0$ otherwise \cite{CE}. Accordingly, the CE minimization problem (\ref{CE2}) can be reformulated as the following probability learning-based optimization problem:
\begin{align} 
\label{CE4}
\max_{\bm{\omega}} &  ~~\frac{1}{{V}_{elite}}\sum_{v=1}^{{V}_{elite}} \ln g(\bm{\beta^{[v]},\bm{\omega}}).
\end{align}

\renewcommand{\algorithmicrequire}{\textbf{Input:}}
\renewcommand{\algorithmicensure}{\textbf{Output:}}
\begin{algorithm}[t] 
	\caption{The Overall Algorithm for solving outer-layer optimization problem (\ref{prob4}) }
	\begin{algorithmic}[1] 
		\State {Initialize  $it=0$ and  $\bm{\omega}^{(it)}=0.5\times \bm{1}$.}
		\Repeat
		\State {Generate $V$ feasible samples satisfying the constraints (\ref{prob29}\text{a}) and (\ref{prob29}\text{b}) based on the Bernoulli distribution function $g(\bm{\beta}, \bm{\omega}^{(it)})$. }
		\State {For each sample, calculate the objective value of the original problem (\ref{prob8}) using \textbf{Algorithm 1}.}
		\State {Sort the $V$ samples in ascending order of objective values.}
		\State {Select the ${V}_{elite}$ samples with the lowest objective values as elite samples.}
		\State {Obtain $\bm{\upsilon}^{(it)}$ by equation (\ref{CE6}).}
		\State {Update $\bm{\omega}^{(it+1)}$ by equation (\ref{CE7}).}	  	    
		\State {Update $it = it + 1$.}
		\Until {The objective value of (\ref{prob4}) converges.}	
	\end{algorithmic}
\end{algorithm}

The optimal probability distribution parameter $\bm{\omega}^*$ is determined as
\begin{equation} \label{CE5}
\bm{\omega}^* = \arg\max_{\bm{\omega}} \frac{1}{{V}_{elite}} \sum_{v=1}^{{V}_{elite}} \ln g(\bm{\beta^{[v]},\bm{\omega}}).
\end{equation}

By substituting (\ref{CE3}) into (\ref{CE5}) and forcing its first-order derivative to zero, the optimal ${\omega}_{kl}$ can be obtained as
\begin{equation} \label{CE6}
{\omega}_{kl}^* = \sum_{v=1}^{{V}_{elite}} \beta_{kl}^{[v]}.
\end{equation}

To update $\bm{\omega}$, we adopt a  smoothing update framework suitable for CE-based algorithms involving discrete variables.  The probability parameter at the $(it+1)$-th iteration is updated as
\begin{equation} \label{CE7}
\bm{\omega}^{(it+1)} = \rho\bm{\upsilon}^{(it)}+(1-\rho)\bm{\omega}^{(it)},
\end{equation}
where $\bm{\upsilon}^{(it)}$ denotes the new probability parameter obtained from (\ref{CE6}), $\bm{\omega}^{(it)}$ represents the probability parameter used for sample generation in the $it$-th iteration,  and $\rho$ is the learning rate.

\subsubsection{Overall Algorithm for the Outer-Layer Optimization  Problem}
To make it clearer, the CE-based algorithm for solving the outer-layer optimization problem is summarized in Algorithm 2.  Specifically, we first generate $V$ feasible samples through random sampling based on the probability distribution parameters $\bm{\omega}^{it}$. For each sample, the inner-layer optimization problem is solved using Algorithm 1, and the samples are then sorted in ascending order of their objective values. The top $V_{elite}$ highest-performing samples are subsequently used to update the probability distribution parameters $\bm{\omega}^{(it+1)}$. This learning process is repeated iteratively until the objective value of problem (\ref{prob4}) converges.

 \subsection{Complexity Analysis} 
 The complexity of solving problem (\ref{prob29}) is primarily from solving the inner-layer optimization problem via Algorithm 1 (\ref{prob23}) under different given samples.  In  Algorithm 1, the per-iteration complexity mainly stems from computing the closed-form solutions of (\ref{MMSE1}), (\ref{MMSE3}), and (\ref{w1}), which involve matrix multiplications and SVD \cite{Liu2018tsp, WMMSE}.  Consequently, the complexity of each iteration for the inner optimization problem is $\mathcal{O}(N_t^3+M^2N_t+N_t^2M)$. Considering the CE-based algorithm for the outer-layer optimization problem, the overall complexity is  $\mathcal{O}(it_Iit_OV(N_t^3+M^2N_t+N_t^2M))$, where $it_I$ and $it_O$ denote the numbers of iterations for Algorithm 1 and Algorithm 2, respectively. It is evident that the proposed CE-based algorithm offers significantly lower complexity compared to the exhaustive searching or BnB approach, whose worst-case complexity is $\mathcal{O}(it_I(\frac{L_k(L_k+1)}{2})^K(N_t^3+M^2N_t+N_t^2M))$.

\section{Simulation Results}

\begin{figure*}[!t]
	\centering
	\includegraphics[width=6in, height=1.4 in]{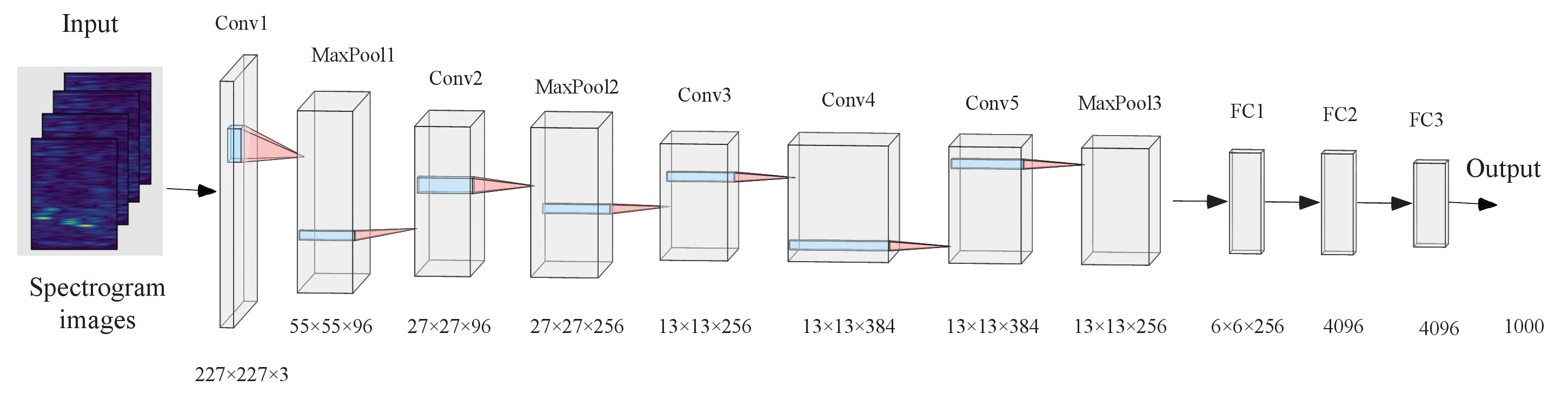}
	\caption{The  AlexNet Architecture.}
	\label{fig:alexnet}
\end{figure*}

In this section, we evaluate the effectiveness of the proposed three-tier DNN partitioning scheme. We first describe the system setup and baseline schemes, followed by a detailed discussion on simulation results.

 \subsection{System Setup and Benchmarks}
Unless otherwise specified, the following system settings are adopted. The BS is located at $(0\,\text{m}, 0\,\text{m})$, and 5 ISAC devices are randomly distributed within a square area of $[-200\,\text{m}, 200\,\text{m}] \times [-200\,\text{m}, 200\,\text{m}]$. The classical AlexNet is  employed to perform the sensing task of target classification~\cite{Alexnet1, Alexnet2}, and its architecture and relevant parameters are shown in Fig.~\ref{fig:alexnet}. In particular, AlexNet consists of five convolutional layers, three pooling layers, and three fully connected layers, and takes an input of size $227 \times 227 \times 3$. For the parameters in Algorithm 2, we set the number of samples $V$ per iteration to 1000, the number of elites $V_{elite}$ to 50, and the learning rate $\rho$ to 0.9. Other key parameters are summarized in Table~I. To ensure statistical reliability, we execute Monte Carlo simulations, with results averaged over 200 independent trials. To validate the superiority of the proposed scheme, we compare it against the following three benchmark schemes under the same configuration.

\begin{table}[t]
	\renewcommand{\arraystretch}{1.25}
	\caption{SIMULATION PARAMETERS.}
	\label{table_example2}
	\centering
	\begin{tabular}{l l}
		\hline
		
		\bfseries Parameters &  \multicolumn{1}{c}{\bfseries Value}\\ 
		\hline
		Number of BS's antennas $M$ & 12\\
		Number of ISAC device's antennas $N_t$  & 8\\
		Transmit signal bandwidth $B$  & 5 MHz  \\
		\tabincell{l}{Devices' maximum transmit power $P_{t}$} & 30 dBm\\
		Noise power spectral density at the BS $\sigma_b^2$ & -174 dBm/Hz\\
		FLOPs per CPU cycle at ISAC device $k$ $\alpha^L_k$ & 2 FLOPs/cycle \\
	    FLOPs per CPU cycle at the MEC server $\alpha^M_k$ & 4 FLOPs/cycle \\
		FLOPs per CPU cycle at cloud server $\alpha^C_k$ & 8 FLOPs/cycle \\
		Computational capacity of ISAC device $k$ $F_k$ & 0.8 Gcycles/s\\
		Computational capacity of MEC server $F_M$ & 12 Gcycles/s\\
		\tabincell{l}{Maximum CPU frequency allocated to ISAC device $k$ \\ by the cloud server $f^C$}  & 20 Gcycles/s\\
			\tabincell{l}{Backhaul rate between MEC server and cloud \\ server $r_b$}  & 2 Mbit/s\\
			The hardware-related power coefficient $\kappa$ & $10^{-28}$ \\
			Computational energy budget of ISAC device  $E_{th}$ & 300 Joule \\
		\hline
		
	\end{tabular}
\end{table}

\begin{itemize}
	\item {\bfseries Local Execution Scheme}: Each device performs the entire DNN inference task locally without any offloading.
	
	\item {\bfseries Edge-Device Two-Tier DNN Partitioning Scheme (ED-DP)~\cite{dnniscc3}}: In this scheme, the DNN is partitioned into two parts: one part is executed locally on the device, while the other is offloaded to BS and processed by the MEC server.
	
	\item {\bfseries Cloud-Edge-Device Three-Tier Scheme without DNN Partitioning (CED-WDP)~\cite{liu2024joint}}: In this  three-tier ISCC framework, the DNN model is executed as a whole without being partitioned. Depending on the specific deployment strategy, the entire DNN model runs either on the cloud server, on the MEC server, or locally.
\end{itemize}

 \subsection{Convergence Performance}

\begin{figure}
	\setlength{\abovecaptionskip}{-0.1 cm}
	\centering
	\begin{subfigure}[]
		{\centering
			\includegraphics[width=1.8in]{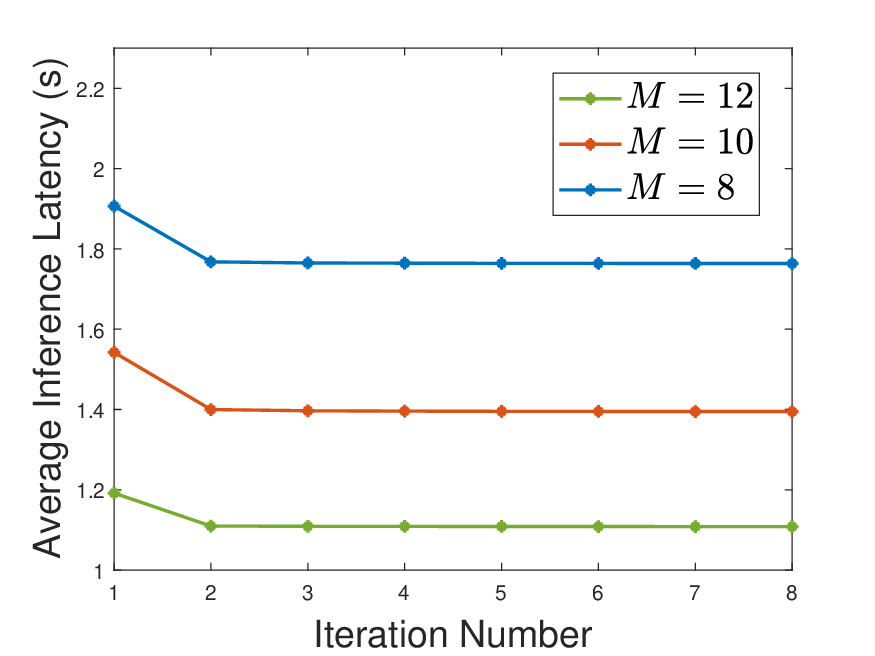}}
	\end{subfigure}
	\hspace{-7 mm}
	\begin{subfigure}[]
		{\centering
			\includegraphics[width=1.8in]{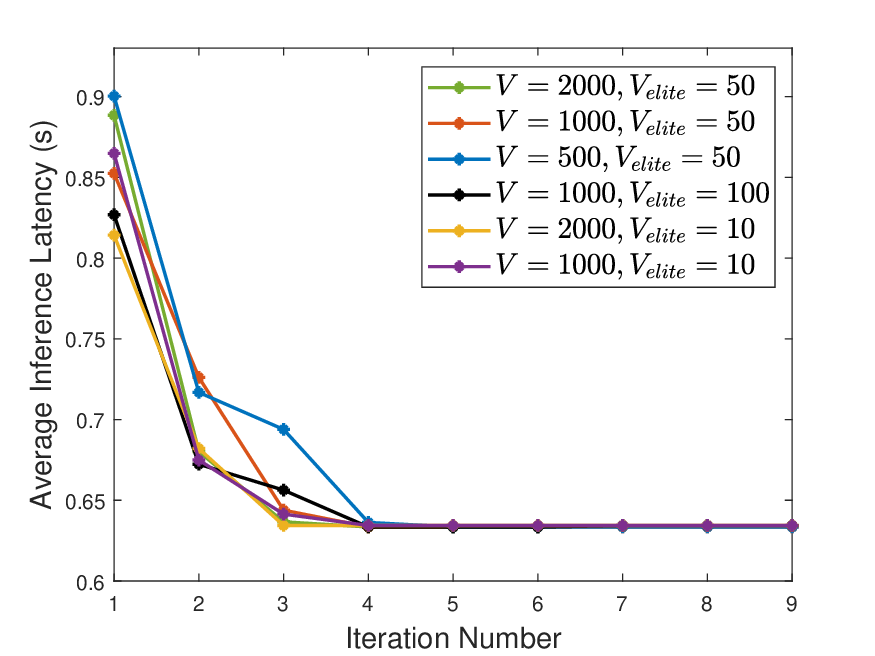}}
	\end{subfigure}
	\caption{(a) The convergence performance of Algorithm 1. (b) The convergence performance of Algorithm 2. }
	\label{fig:f1}
\end{figure}

In Fig.~\ref{fig:f1}(a) and Fig.~\ref{fig:f1}(b), we present the convergence performance of the proposed Algorithm~1 and Algorithm~2, respectively. It is observed that both algorithms converge \textcolor{black}{quickly} within a finite number of iterations. For Algorithm~1, the average inference latency decreases as the number of BS antennas \( M \) increases. This is because additional receive antennas enable the BS to more effectively suppress multi-user interference through receive beamforming, thereby improving the offloading rate. Furthermore, we observe that the convergence behavior of Algorithm~2 is affected by both number of samples \( V \) and elite samples \( V_{\text{elite}} \). In general, larger \( V \) and smaller \( V_{\text{elite}} \)  lead to faster convergence.

 \subsection{Average Inference Latency and BeamPattern Performance}
\begin{figure}[t]
	\centering
	\includegraphics[width=3.2in]{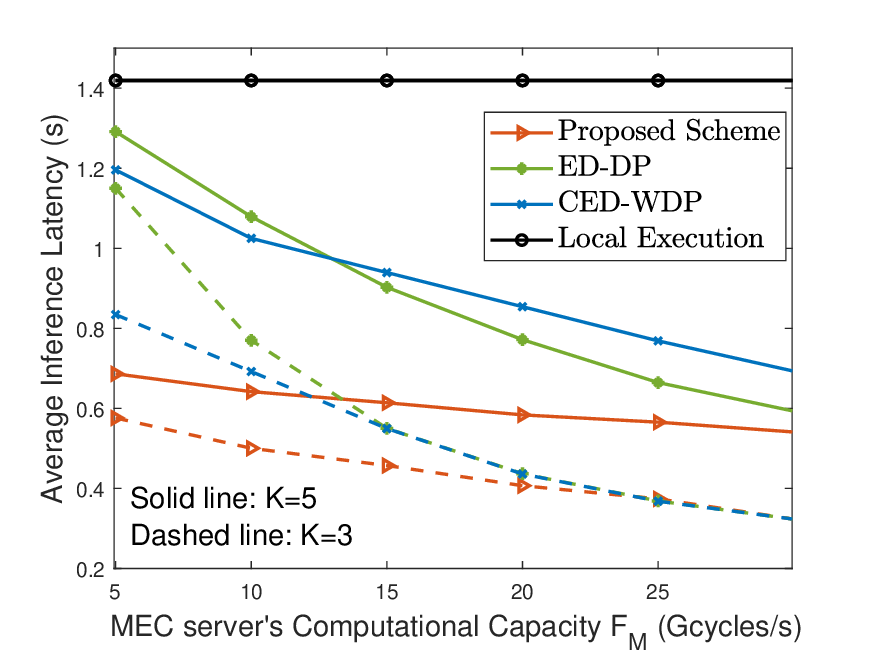}
	\caption{Average inference latency versus different MEC server's computational capacity $F_M$.}
	\label{fig:f2}
\end{figure}

Fig.~\ref{fig:f2} illustrates the average inference latency of different schemes under varying MEC server computational capacity $F_M$. It can be observed that, except for the local execution scheme, the inference latency of all other schemes decreases as $F_M$ increases, due to the availability of more edge computational resources. When $F_M$ is small, CED-WDP achieves lower latency than ED-DP, as it can offload tasks to the cloud server with higher computational power when MEC resources are limited.  However, as $F_M$ increases, ED-DP outperforms CED-WDP due to the flexibility enabled by DNN partitioning, which allows adaptive selection of partition points to reduce fronthaul load and enhance resource utilization. By integrating the advantages of both approaches, the proposed scheme achieves the lowest average inference latency across all configurations. Moreover, unlike ED-DP and CED-WDP, the proposed scheme exhibits greater robustness to variations in $F_M$, \textcolor{black}{acting as lower bound for ED-DP and CED-WDP schemes,} owing to its flexible partitioning strategy and adaptive offloading mechanism. In scenarios with limited MEC capacity, the proposed framework is able to maintain stable latency by dynamically reallocating computational tasks between the cloud server and devices.

\begin{figure}[t]
	\centering
	\includegraphics[width=3.2in]{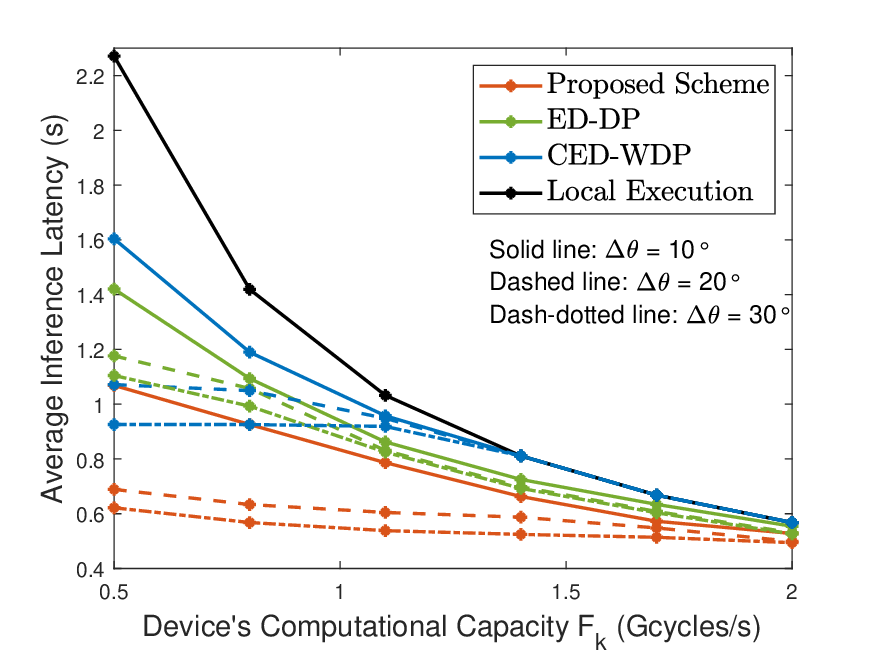}
	\caption{Average inference latency versus different ISAC device's Computational Capacity $F_k$.}
	\label{fig:f3}
\end{figure}

Fig.~\ref{fig:f2} also illustrates the impact of the number of ISAC devices $K$ on the average inference latency across different schemes. As $K$ decreases, the latency under ED-DP, CED-WDP, and the proposed scheme consistently declines. This trend is primarily attributed to two factors: (i) fewer devices allow more computational resources to be allocated per device at the MEC server, and (ii) reduced inter-device interference improves the task offloading rate. Moreover, we observe that when $K=3$ and the MEC server has sufficient computational capacity, all three schemes yield the same average inference latency. This is because, under such conditions, the optimal strategy for all schemes is to offload the entire tasks  to the MEC server.

Fig.~\ref{fig:f3} presents the average inference latency under different ISAC device computational capacities $F_k$. As $F_k$ increases, the average inference latency of all schemes decreases and gradually converges. This trend is attributed to the diminishing benefits of cloud processing and DNN partitioning, as enhanced local computational capability reduces the need for offloading. Moreover, Fig.~\ref{fig:f3} also illustrates the impact of the beampattern mainlobe width $\Delta\theta$ on latency performance. It can be observed that increasing $\Delta\theta$ results in reduced inference latency for CED-WDP, ED-DP, and the proposed scheme.  This is because a wider mainlobe enables the beam energy to more effectively cover communication targets and scattering paths, thereby improving the task offloading rate.

\begin{figure}[t]
	\centering
	\includegraphics[width=3.2in]{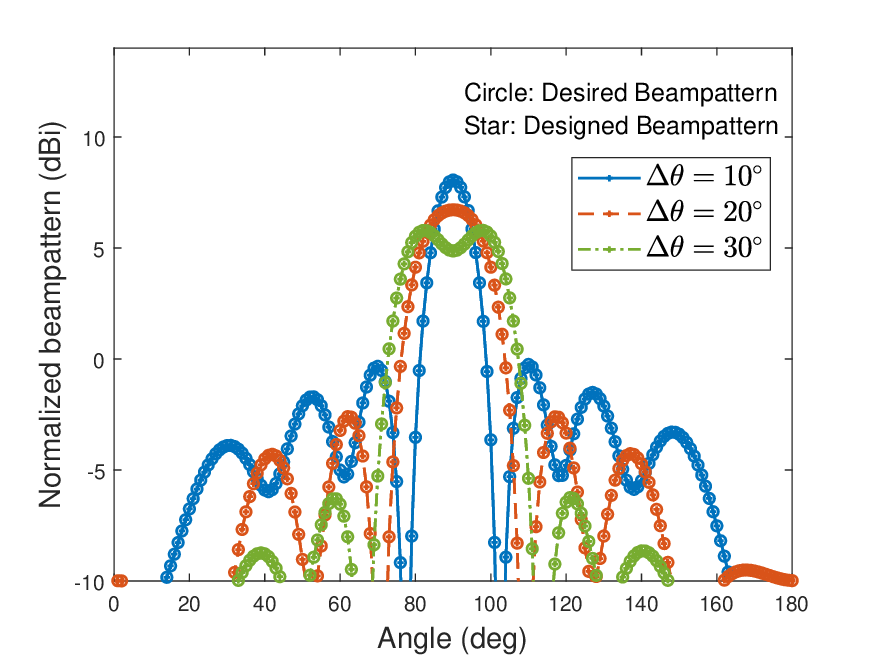}
	\caption{Beampattern versus the mainlobe width $\Delta\theta$.}
	\label{fig:f4}
\end{figure}

In Fig.~\ref{fig:f4}, we present the beampatterns corresponding to different mainlobe widths $\Delta\theta$. It can be observed that a smaller $\Delta\theta$ yields a higher beamforming gain, which is more favorable for target sensing. Moreover, due to the equality constraint (\ref{prob29}f), the beampattern generated by the designed ISAC beamforming matrix precisely matches the desired beampattern obtained by solving the least-squares problem (19) in \cite{beam}, thereby validating the effectiveness of the proposed algorithm in preserving sensing performance.

 \begin{figure}[t]
 	\centering
 	\includegraphics[width=3.2in]{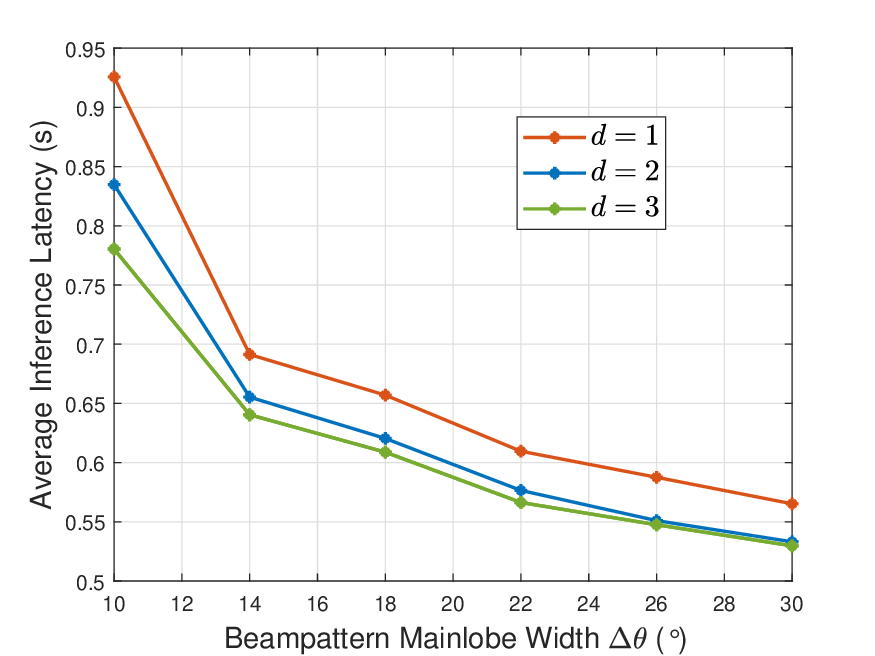}
 	\caption{Average inference latency versus beampattern mainlobe width $\Delta\theta$.}
 	\label{fig:f5}
 \end{figure}
 
To further illustrate the trade-off between sensing and computation, Fig.~\ref{fig:f5} shows the variation in average inference latency with respect to the beampattern  mainlobe width $\Delta\theta$. It can be observed that the average inference latency decreases as $\Delta\theta$ increases. Meanwhile, as shown in Fig.~\ref{fig:f4}, a larger $\Delta\theta$ corresponds to a lower beamforming gain, highlighting the inherent trade-off between computational efficiency and sensing performance. Moreover, as the number of streams $d$ increases, a more favorable latency–mainlobe width trade-off can be achieved due to the improved offloading rate.

 \subsection{Optimality and Complexity Analysis}
 \begin{figure}[tbp]
	\centering
	\includegraphics[width=3.2in]{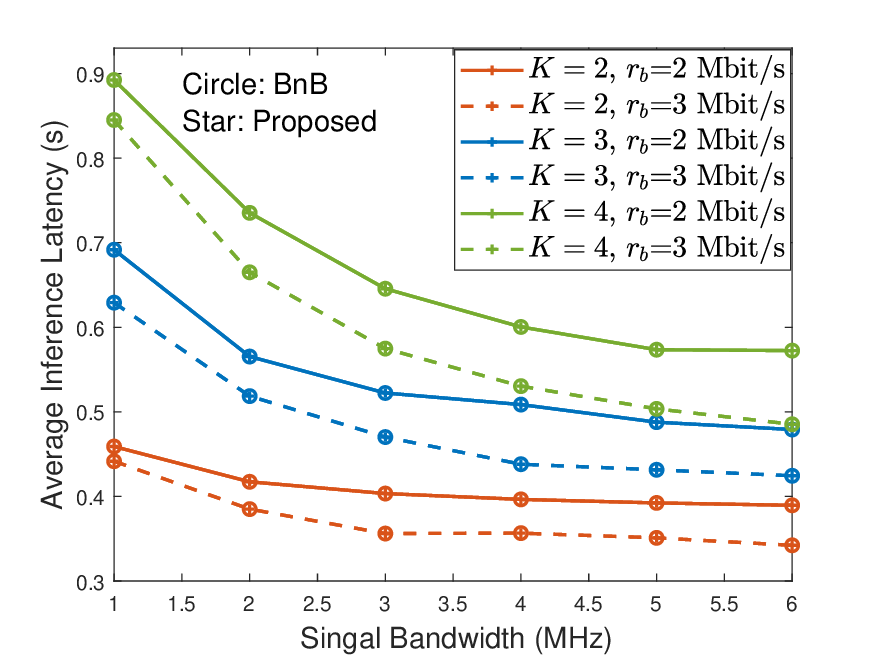}
	\caption{Average inference latency versus signal bandwidth  $B$.}
	\label{fig:f6}
\end{figure}
\renewcommand{\arraystretch}{1.35}  
\begin{table} [t]
	
	\centering
	\caption{The  CPU runtime of two algorithms under different numbers of devices $K$ }
	\label{Tbl:Suspiciousness}
	\begin{tabular}{|c|c|c|c|}
		\hline
		\setlength{\abovecaptionskip}{0 cm}
		\setlength{\belowcaptionskip}{0cm}
		\multirow{2}{*}{\tabincell{c}{Number \\  of the devices}}  &\multirow{2}{*}{\tabincell{c}{The BnB  \\ Algorithm}} &\multirow{2}{*}{\tabincell{c}{Proposed \\ Algorithm} }  \\& & 
		\\ \hline
		
		$K=2$     & 11.7321 s & 6.402 s  \\  \hline 
		$K=3$     & 646.067 s & 15.024 s   \\  \hline 
		$K=4$    & 27716.613 s & 19.407 s   \\ 
		\hline
		
	\end{tabular}
\end{table}
\textcolor{black}{Although the formulated problem can hardly be optimally solved, especially for a large-scale network, we show the effectiveness} of the proposed CE-based algorithm in designing DNN partitioning strategies in Fig.~\ref{fig:f6}, where the achieved average inference latency is compared  with that of the BnB algorithm, which provides the optimal solution, under different bandwidths $B$, numbers of ISAC devices $K$, and MEC server-to-cloud backhaul rates $r_b$. It is observed that the proposed algorithm achieves identical performance to the BnB algorithm across all settings. Moreover, as $B$ and $r_b$ increase, the average inference latency decreases accordingly, due to the reduction in transmission delay of intermediate features. Furthermore, Table~\ref{Tbl:Suspiciousness} shows the average runtime of the two algorithms under different $K$, evaluated using MATLAB R2022 on an Intel Core i9-13900K processor. As shown, the runtime of the BnB algorithm increases exponentially with $K$, while that of the proposed algorithm grows only modestly. These results further demonstrate the computational efficiency of the proposed algorithm.

\section{Conclusion}
In this paper, \textcolor{black}{we have} proposed a three-tier collaborative inference framework for ISCC networks, in which the cloud server, MEC servers, and ISAC devices cooperatively execute different segments of a pre-trained DNN to perform intelligent sensing tasks. By enabling ISAC-based intermediate feature offloading and leveraging MIMO technology, the framework effectively reduces fronthaul load and improves both sensing and offloading performance. To minimize the overall inference latency, we formulated a joint optimization problem involving DNN partitioning, ISAC beamforming, and computational resource allocation, subject to sensing constraints. \textcolor{black}{We have developed an} efficient two-layer optimization algorithm to solve the joint problem. For the inner layer, \textcolor{black}{we have} derived closed-form solutions using KKT conditions for computational resource allocation, and \textcolor{black}{we have used} an MM-WMMSE-based method to optimize ISAC beamforming. For the outer layer, \textcolor{black}{we have employed} a CE-based learning algorithm  to iteratively update the DNN partitioning strategy.  Simulation results \textcolor{black}{have demonstrated} that the proposed framework significantly reduces inference latency compared to existing two-tier schemes and reveals a clear trade-off between sensing performance and computational efficiency. Furthermore, the algorithm achieves near-optimal performance with substantially lower complexity, making it well-suited for practical deployment in future intelligent ISCC systems.


\appendix

\bibliographystyle{IEEEtran}
\bibliography{biblp/bibfilelp2}

\end{document}